\documentclass[amssymb,aps,twocolumn,floats]{revtex4}
\usepackage{color,graphicx,pstricks}

\begin{document}

\title{ Disorder-induced orbital ordering in doped manganites}

\author{Sanjeev Kumar$^{1,2}$ and Arno P. Kampf$^{3}$ }

\affiliation{
$^{1}$ Faculty of Science and Technology, University of Twente, P.O. Box 217, 7500 AE Enschede, The Netherlands \\
$^{2}$ Instituut-Lorentz for Theoretical Physics, Leiden University,
P.O. Box 9506, 2300 RA Leiden, The Netherlands \\
$^3$ Theoretical Physics III, Center for Electronic
Correlations and Magnetism, Institute of Physics, \\
University of Augsburg, D-86135 Augsburg, Germany
}

\begin{abstract}

We study the effect of quenched disorder on the ordering of orbital and
magnetic degrees of freedom in a two-dimensional, two-band double-exchange 
model for $e_g$ electrons coupled to Jahn-Teller distortions. Using a 
real-space Monte Carlo method, we find that disorder can
induce a short-range ordering of the orbital degrees of freedom
near 30\% hole doping. The most striking consequence of this short range
ordering is a strong increase in the low temperature resistivity.
The real-space approach allows to analyze the spatial patterns of the
charge, orbital, and magnetic degrees of freedom, and the correlations among 
them. The magnetism is inhomogeneous on the nanoscale in the short-range 
orbitally ordered state.

\vskip 0.2cm


\vskip 0.2cm


\end{abstract}

\maketitle

\section{Introduction}

Hole-doped perovskite manganites RE$_{1-x}$AE$_x$MnO$_3$ (RE=rare earth, 
AE=alkaline earth) have attracted great attention from the condensed matter 
community over the last decade \cite{dagotto_book}. While the initial surge of 
research activities on these materials was triggered by the discovery of the 
colossal magnetoresistance (CMR) effect, a rich variety of phases and phase transitions was subsequently uncovered \cite{tapan_book,tokura_book}. 
It is now widely accepted that the interplay among charge, spin, orbital and lattice degrees of freedom
is the underlying cause of the complexity and richness of the physical phenomena observed in manganites.
Recent efforts from both experiment and theory have highlighted the significance of
quenched disorder in these materials \cite{attfield-98,akahoshi,tokura-gpd,attfield-01}.
Therefore, analyzing the effects of disorder in manganites has become an active
area of research \cite{dag2,moreo,sk-apk-pm,kalpa-PRL}.

Disorder is generally viewed as an agent for suppressing the
ordering tendencies of the microscopic degrees of freedom.
Experiments on the half-doped ($x=0.5$) manganites show
that quenched disorder indeed spoils the long-range ordering of the
charge, orbital, and spin variables leading, in some cases, to a
short-range ordering of these microscopic degrees of freedom
\cite{akahoshi,tokura-gpd}. The opposite effect, however, is observed in manganites near
30\% hole doping, where an ordering of the orbital degrees of freedom is
induced by the presence of quenched disorder \cite{attfield-01}.

   In manganites, the average $r_A$ and the variance 
$\sigma^2$ of the A-site ionic radii are known to control the single-particle
bandwidth and the magnitude of quenched disorder, respectively \cite{note1}.
Samples with constant $r_A$ and varying $\sigma^2$ were used in the experiments
of Ref. \cite{attfield-01} with a combination of La, Pr, Nd, Sm, and Ca, Sr, Ba at the 
A-site, while keeping $x=0.3$.
An increase in the low-temperature resistivity by four orders of magnitude
was attributed to the onset of orbital ordering, which was also evidenced from 
the structural changes analyzed via powder x-ray diffraction. Magnetism is 
affected strongly with a reduction in both, the Curie temperature $T_C$ and 
the saturation value of the magnetization. This doping regime is also believed 
to be magnetically inhomogeneous, as independently inferred from NMR and 
neutron scattering experiments \cite{NMR,neut-scatt}.

Disorder has been included previously in models for manganites to study its
influence on the long-range ordered phases \cite {dag2,motome,dag1} especially 
near a first-order phase boundary or in the vicinity of phase separation
\cite{moreo,kalpa-PRL,nano-PRL}. The idea that quenched disorder may lead to 
a partial ordering of the orbital degrees of freedom in manganite models has 
so far remained unexplored.

In this paper, we study a two-band double-exchange model with quenched 
disorder using a real-space Monte Carlo method.
Disorder is modelled via random on-site energies selected from a given 
distribution. We consider two different types of distributions, described in 
detail in the next section. Here and below we refer to these distributions as 
(i) binary disorder and (ii) random scatterers.
While the binary disorder has no significant effect on the orbital degrees of 
freedom, random scatterers lead to orbitally ordered (OO) regions, and
a sharp increase in the low-temperature resistivity is found as observed in the
experiment \cite{attfield-01}.
The magnetic structure is inhomogeneous in a restricted doping regime as 
observed in the NMR and the neutron scattering experiments. Within clusters, 
staggered orbital ordering is accompanied by ferromagnetism, thus providing an example of 
Goodenough-Kanamori rules in an inhomogeneous system \cite{goodenough-kanamori,Imada}.

\section{Model and Method }
We consider a two-band model for itinerant $e_g$ electrons on a square lattice.
The electrons are coupled to Jahn-Teller (JT) lattice distortions,
$t_{2g}$-derived $S = 3/2$ core spins and quenched disorder as described by 
the Hamiltonian:
\begin{eqnarray}
H &=& \sum_{\langle ij \rangle \sigma}^{\alpha \beta}
t_{\alpha \beta}^{ij} ~
c^{\dagger}_{i \alpha \sigma} c^{~}_{j \beta \sigma}
+ \sum_i \epsilon_i ~ n_i + J_s\sum_{\langle ij \rangle} {\bf S}_i \cdot {\bf S}_j \cr
&&
~ - J_H\sum_i {\bf S}_i.{\mbox {\boldmath $\sigma$}}_i 
 - \lambda \sum_i {\bf Q}_i.{\mbox {\boldmath $\tau$}}_i 
+ {K \over 2} \sum_i {\bf Q}_i^2. ~ ~ ~ ~ ~ ~ ~ ~ ~
\end{eqnarray}
\noindent
Here, $c$ and $c^{\dagger}$ are annihilation and creation operators for
$e_g$ electrons, $\sigma = \uparrow,\downarrow$ is the spin index and
$\alpha$, $\beta $ are summed over the two Mn-$e_g$ orbitals
$d_{x^2-y^2}$ and $d_{3z^2-r^2}$, which are labelled $(a)$ and $(b)$ in what follows.
$t_{\alpha \beta}^{ij}$ denote the hopping amplitudes between
$e_g$ orbitals on nearest-neighbor sites and have the cubic perovskite specific form:
$t_{a a}^x= t_{a a}^y \equiv t$,
$t_{b b}^x= t_{b b}^y \equiv t/3 $,
$t_{a b}^x= t_{b a}^x \equiv -t/\sqrt{3} $,
$t_{a b}^y= t_{b a}^y \equiv t/\sqrt{3} $, where
$x$ and $y$ mark the spatial directions \cite{dagotto99}.
The $e_g$-electron spin is locally coupled to the
$t_{2g}$ spin ${\bf S}_i$ via the Hund's rule coupling 
$J_H$.
The $e_g$-electron spin is given by ${\sigma}^{\mu}_i= 
\sum_{\sigma \sigma'}^{\alpha} c^{\dagger}_{i\alpha \sigma} 
\Gamma^{\mu}_{\sigma \sigma'}
c_{i \alpha \sigma'}$,
where $\Gamma^{\mu}$ are the Pauli matrices.
$J_s$ is the strength of the superexchange coupling between neighboring 
$t_{2g}$ spins. $\lambda$ denotes the strength of the JT coupling between the 
distortion ${\bf Q}_i=(Q_{ix},Q_{iz})$ and the orbital pseudospin
${\tau}^{\mu}_i=\sum^{\alpha\beta}_{\sigma}c^{\dagger}_{i\alpha \sigma} 
\Gamma^{\mu}_{\alpha \beta} c_{i\beta \sigma}$.
$K$ is a measure of the lattice stiffness, and we set $t=1=K$ as our 
reference energy scale.

The following two forms of on-site disorder modelling are used:
(i) binary disorder: $\epsilon_i$ takes equally probable values $\pm \Delta$,
(ii) random scatterers: a fraction $x$ of the sites are taken to have 
$\epsilon_i=D$, while for the other sites $\epsilon_i=0$. Although the first 
choice of disorder is the simplest from the model point of view, the second 
appears more realistic. In real materials a fraction $x$ of the rare-earth 
ions is replaced by alkaline-earth ions at random locations. Therefore, it is 
likely that the disorder arising as a consequence of this substitution is 
connected to the amount of doping. This situation is modelled by placing 
repulsive potentials on a fraction $x$ of the sites, which are randomly 
selected. A typical measure of the strength of a disorder distribution is its
variance. For the binary distribution the variance is $\Delta$, while for the
finite density $x$ of scatterers with potential strength $D$, it is 
$D\sqrt{x(1-x)}$. These two models for disorder were previously employed
in a study of half-doped manganites \cite{kalpa-PRL}. The JT distortions and 
the $t_{2g}$ derived core spins are treated as classical variables, and we set 
$\vert {\bf S}_i \vert =1$. Guided by earlier estimates for the JT coupling strength in manganites,
we fix $\lambda = 1.5$ \cite{EP-coupling}, and explore the variation in the parameters $\Delta$, $D$, and
$J_s$.

We further adopt the simplifying limit $J_H >> t $, which is justified and 
frequently used in the context of manganites \cite {moreo, dagotto99, brey}.
In this limit the electronic spin at site $i$ is tied to the orientation of 
the core spin ${\bf S}_i$. Transforming the fermionic operators to this local 
spin reference frame leads to the following effectively 'spinless' model for 
the $e_g$ electrons:
\begin{eqnarray}
H &=& \sum_{\langle ij \rangle }^{\alpha \beta}
{\tilde t}_{\alpha \beta}^{~ij}
 c^{\dagger}_{i \alpha } c^{~}_{j \beta }
+ \sum_i \epsilon_i n_i + J_s\sum_{\langle ij \rangle} {\bf S}_i \cdot 
{\bf S}_j \cr
&& ~ ~ ~ ~
 - \lambda \sum_i {\bf Q}_i.{\mbox {\boldmath $\tau$}}_i 
+ {K \over 2} \sum_i {\bf Q}_i^2.
\end{eqnarray}
The new hopping amplitudes ${\tilde t}$ have an additional dependence on the 
core-spin configurations and are given by:
\begin{eqnarray}
\frac{\tilde t_{\alpha \beta}}{t_{\alpha \beta}} = \cos \frac{\theta_i}{2}
\cos \frac{\theta_j}{2}
+\sin \frac{\theta_i}{2} \sin \frac{\theta_j}{2} ~ e^{-{\rm i}(\phi_i-\phi_j)}.
\end{eqnarray}
\noindent
Here, $\theta_i$ and $\phi_i$ denote polar and azimuthal angles for the spin 
${\bf S}_i$. From now on the operator $c_{i \alpha}$ ($c^{\dagger}_{i \alpha}$)
is associated with annihilating (creating) an electron at site $i$ in the 
orbital $\alpha$ with spin parallel to~${\bf S}_i$.

The model given by Eq. (2) is bilinear in the electronic operators and does 
not encounter the problem of an exponentially growing Hilbert space, since all 
many-particle states can be constructed from Slater determinants of the
single-particle states. The difficulty, however, arises from the large phase 
space in the classical variables ${\bf Q}$ and ${\bf S}$. Exact 
diagonalization based Monte Carlo (ED-MC) is a numerically exact method to 
treat such problems, and has been used extensively in the past 
\cite{moreo,dagotto99,brey}. The classical variables are sampled by the 
Metropolis algorithm, which requires the exact eigenenergy spectrum. Therefore 
iterative ED of the Hamiltonian is needed, which leads to $N^4$ scaling of the 
required cpu time, $N$ is the number of lattice sites. The $N^4$ scaling makes 
this method very restrictive in terms of the achievable lattice sizes, with 
the typical size in previous studies being $\sim 100$ sites. Since a study of 
larger lattices is essential for analyzing the nature of inhomogeneities in 
manganite models, several attempts have been made to devise accurate 
approximate schemes \cite{hybridMC, furukawa-dag, tca}. In the present study 
we employ the travelling cluster approximation (TCA) \cite{tca}, which indeed 
has been very successful in analyzing similar models in the recent past
\cite{dhde,kalpa-PRL, sk-apk-pm}.

\section{Results and Discussion}
We begin with results for bulk quantities describing the ordering of the
magnetic and the lattice degrees of freedom. We focus on the 30\%
hole-doped system
($x=0.3$) for a close correspondence to the experiments in Ref. \cite{attfield-01}.
Fig. 1(a) shows the
effect of binary disorder on the temperature dependence of the magnetization
$m$, defined via $m^2=\langle ( N^{-1} \sum {\bf S}_i )^2 \rangle_{av}$.
Here and below $\langle ... \rangle_{av}$ denotes the average over thermal 
equilibrium configutations and additionally over realizations of quenched 
disorder. Results for disordered systems are averaged over $4-6$ realizations 
of disorder. Clearly, the magnetism is not affected much by the presence of 
weak binary disorder. This is in agreement with previous studies, which find 
that the reduction in $T_C$ is proportional to $\Delta^2$ for weak disorder 
\cite{salafranca, motome1, bouzerar}.

Fig. 1(b) shows the temperature dependence of the ${\bf q}={\bf q}_0\equiv 
(\pi,\pi)$ component of the lattice structure factor 
$D_Q({\bf q})=N^{-2}\sum_{ij}\langle {\bf Q}_i\cdot{\bf Q}_j\rangle_{av}~
e^{-{\rm i}{\bf q}\cdot ({\bf r}_i-{\bf r}_j)}$. $D_Q({\bf q}_0)$ is a measure 
for the staggered distortion order in the system. The lattice ordering leads
to orbital ordering via the JT coupling. An increase with $\Delta$ in the 
low-temperature value of $D_Q({\bf q}_0)$ suggests the appearance
of orbital order. However, this effect is too weak to explain the experimental 
resistivity data \cite{attfield-01}. Moreover, the increase at low $T$ in 
$D_Q({\bf q}_0)$ is not monotonic, which becomes clear by comparing the 
results for $\Delta = 0.4$, $0.8$ and $1.0$ in Fig. 1(b).

\begin{figure}
\centerline{
\includegraphics[width=8.8cm , clip=true]{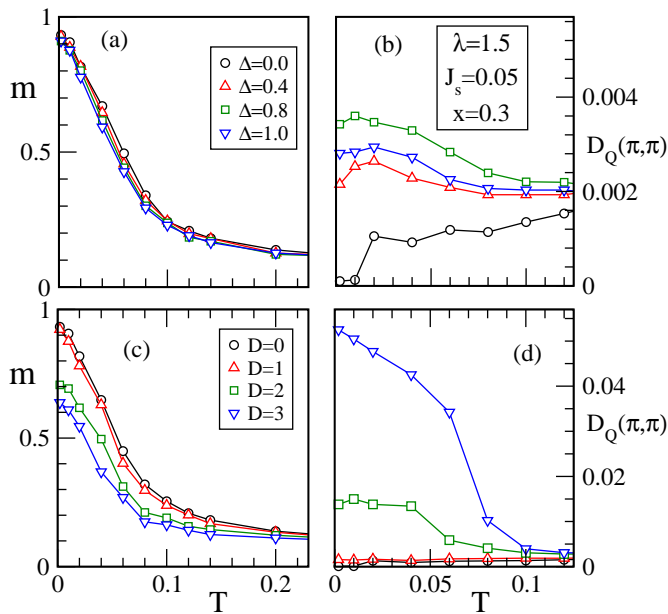}
}
\vspace{.1cm}
\caption{(Color online)
Temperature dependence of (a) the magnetization $m$ and (b) the lattice 
structure factor $D_Q({\bf q})$ at ${\bf q}=(\pi,\pi)$ for various values of 
the disorder strength $\Delta$. (c) and (d) show the same quantities as in (a) 
and (b), respectively, if the on-site disorder is modelled by random 
scatterers of strength $D$. The concentration of scatterers is equal to the 
hole density $x=0.3$. Note the order of magnitude difference in the magnitudes 
for $D_Q({\bf q}_0)$ between panels (b) and (d). All results are at 
$\lambda=1.5$ and $J_s=0.05$.
}
\end{figure}

Now we explore the results for the disorder arising from random scatterers of 
strength $D$. Since the disorder originates from the replacement of RE$^{3+}$ by
AE$^{2+}$ ions, the 
density of random scatterers is kept equal to the doping concentration $x$.
$m(T)$, shown in Fig. 1(c), is affected strongly upon increasing $D$, with a 
decrease in the saturation value of the magnetization pointing towards a
magnetically inhomogeneous groundstate. More importantly, a monotonic increase 
with $D$ is observed in the low-temperature values of $D_Q({\bf q}_0)$ (see 
Fig. 1(d)). The rise in $D_Q({\bf q}_0)$ clearly indicates the emergence of 
orbital ordering in the system, with the area and/or strength of the ordered 
regions increasing with increasing $D$.

\begin{figure}
\centerline{
\includegraphics[width=8.8cm , clip=true]{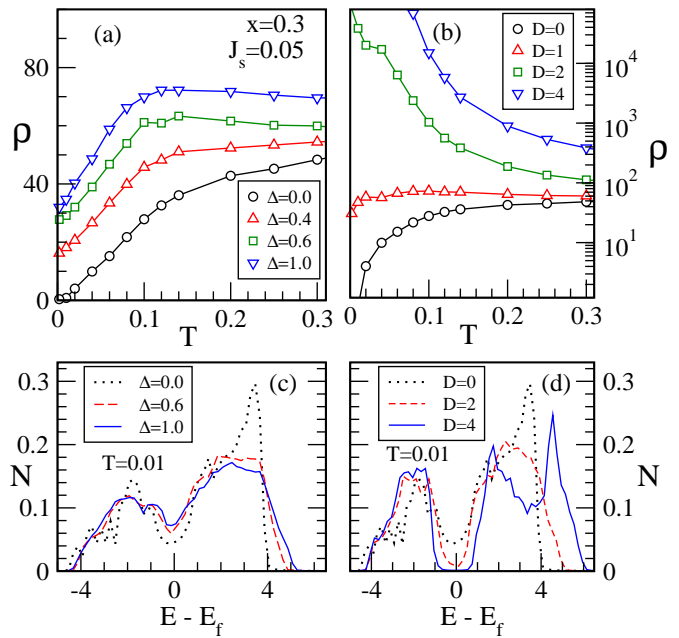}
}
\vspace{.1cm}
\caption{(Color online)
Temperature dependent resistivity $\rho(T)$ (in units of $\hbar /\pi e^2$) for 
varying strength of (a) binary disorder $\Delta$, and (b) random scatterers 
$D$. Note the logarithmic scale in (b). (c)-(d) Low-temperature density of 
states for the two types of disorder.
}
\end{figure}

It is expected that these orbital ordering correlations are reflected in the transport
properties. We therefore compute the dc resistivity $\rho$ approximated by the
inverse of low-frequency optical conductivity, which is calculated by using
exact eigenstates and -energies in the Kubo-Greenwood formula \cite{transport}.
Fig. 2(a)-(b) show $\rho$ as a function of temperature for the two disorder 
models described above. The low-temperature resistivity increases upon 
increasing the binary disorder strength $\Delta$ (see Fig. 1(a)). For small 
values of $\Delta$ the resistivity curves appear parallel to each other below 
$T\sim 0.1$. The resistivity therefore follows Mathiessen's rule,
{\it i.e.} $\rho(T)$ for the disordered system is obtained from $\rho(T)$ for 
the clean system by simply adding a constant contribution arising from the
scattering off the disorder potential. $d\rho/dT$ remains positive at low 
temperature indicating a metallic behavior. This oversimplified 
description, however,  does not take into account the disorder induced changes 
in the orbital ordering correlations and the related changes in the density of 
states (discussed below).

\begin{figure}
\centerline{
\includegraphics[width=8.8cm , clip=true]{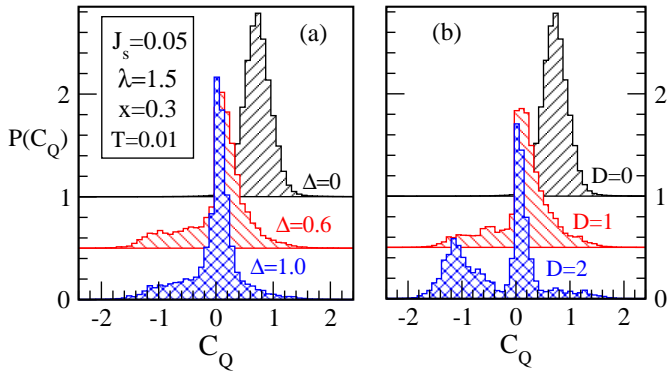}
}
\vspace{.1cm}
\caption{(Color online)
Low-temperature distribution functions generated from the Monte Carlo data for 
the nearest-neighbor correlations $C_Q$ of the lattice distortions for (a) 
binary disorder and (b) for random scatterers. The curves for different 
$\Delta$ are off-set along the y-axis for clarity. $C_Q$ is positive 
(negative) for ferro- (antiferro)-distortive patterns of the lattice variables.
}
\end{figure}

Random scatterers lead to a drastically different behavior. The low-temperature
rise in $\rho(T)$ covers several orders of magnitude (see Fig. 2(b)).
The negative sign of $d\rho/dT$ for $D>1$ signals an insulating behavior. Upon increasing
the disorder strength $D$ we therefore observe a metal to insulator transition.
For $x=0.3$ both disorder models have the same variance, if $\Delta \sim 0.46 D$ holds.
Comparing, therefore, the results for $D=2$ and $\Delta  = 1$, we have to conclude that 
the drastic rise in the resistivity for random scatterers can not be attributed to the strength 
of the disorder potential. In fact, a large increase in the low-$T$ resistivity was 
one of the experimental indications for the onset of disorder-induced 
orbital ordering \cite{attfield-01}. 

Figs. 2(c)-(d) highlight the difference between the densities of states (DOS) 
for the two choices of disorder modelling. The DOS is defined as $N(\omega)=
\langle N^{-1}\sum_i\delta (\omega-E_i)\rangle_{av}$, where $E_i$ denotes the 
eigenvalues of the Hamiltonian. Here we approximate the $\delta$-function by a 
Lorentzian with width $\gamma=0.04$:
\begin{equation}
\delta(\omega -E_i)\simeq \frac{\gamma/\pi}{[\gamma^2+(\omega-E_i)^2]}.
\end{equation}
The DOS for the clean system has a pseudogap structure near the chemical 
potential. For binary disorder, the pseudogap slowly fills up with increasing 
$\Delta$. In contrast, it deepens upon adding random scattering centers and 
even leads to a clean gap for $D\geq 3 $. This opposite behavior is partly 
responsible for the drastically different low-temperature resistivity 
discussed above. The three-peak structure for large values of $D$ in Fig. 2(d) can be 
understood as follows: A fraction $2x$ of electronic states split off and form a narrow impurity band 
centered at an energy $D$ above the Fermi level of the undoped system.
The lower band now contains a fraction $2(1-x)$ of the states with 
the Fermi level located in the middle of the band. This leads to a situation 
similar to the undoped system, and an energy gap originating from staggered orbital 
ordering opens at the Fermi level.

To gain further insight into the nature of the states in the presence of the 
two types of disorder, we plot the distribution functions for the lattice 
variables in Fig. 3. Panel (a) shows the distribution of the nearest neighbor 
lattice correlations, $C_Q(i)=(1/4)\sum_{\delta}{\bf Q}_i\cdot{\bf Q}_{i+\delta}$
for binary disorder; here $\delta$ denotes the four nearest neighbor sites of 
site $i$. A negative value of $C_Q(i)$ indicates an antiferro pattern
of JT distortions, and hence a pattern of staggered orbital ordering. The 
distribution function for $C_Q$ is defined as $P(C_Q)=\langle N^{-1}\sum_i 
\delta (C_Q-C_Q(i))\rangle_{av}$; the $\delta$-function is again approximated 
by a Lorentzian with width $\sim 0.04$. A peak in $P(C_Q)$ centered near $C_Q 
=0.8$ for $\Delta=0$ indicates that the clean system has weak 
ferro-distortive/ferro-orbital correlations. Tails going down to $C_Q\sim -1.4$
arise in the distribution function upon including binary disorder.

\begin{figure}
\centerline{
\includegraphics[width=8.6cm , clip=true]{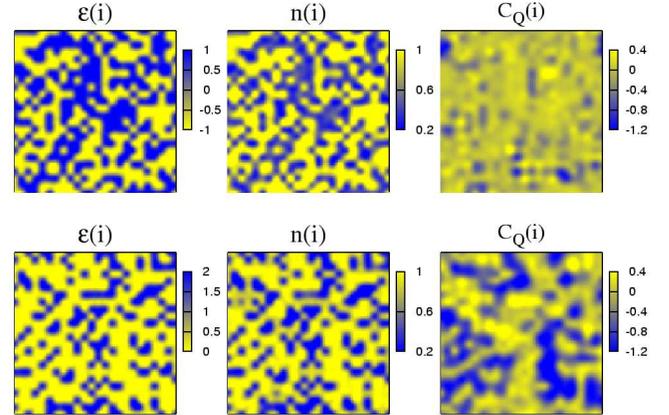}
}
\vspace{.1cm}
\caption{(Color online)
Real-space patterns of the disorder potential $\epsilon(i)$, charge density 
$n(i)$, and lattice correlations  $C_Q(i)$. Top row: binary disorder with 
$\Delta =1$, bottom row: random scatterers with $D=2$. The patterns in both 
cases are shown on a $24\times 24$ lattice for a single disorder realization 
at $T=0.01$ and $x=0.3$.
}
\end{figure}

The distribution function $P(C_Q)$ for random scatterers looks qualitatively 
different. We recall that the strengths of the two types of disorder are 
related via $\Delta\sim 0.46 D$. The low-temperature distributions $P(C_Q)$ are
plotted in Fig. 3(b) for random scatterers. A qualitative change in the shape 
of the distribution function occurs for $D=2$, where a second peak centered 
around $C_Q\sim -1.2$ emerges. This is a direct indication that a significant 
fraction of the system becomes orbitally ordered. This correlates perfectly 
with the strong rise in $D_Q({\bf q}_0)$ at low temperatures 
(see Fig. 1(b)), and the anomalous increase in the resistivity (see Fig. 2(b)).

A real-space picture for the emergence of orbital ordering is presented in 
Fig. 4, which displays the disorder potential $\epsilon_i$, the electronic 
density $n_i$, and the lattice correlations $C_Q(i)$. The top row for binary 
disorder shows that the charge density closely follows the disorder potential.
The local lattice correlations are centered around $C_Q = 0$, which is also 
evident from the peak in the distribution $P(C_Q)$ shown in Fig. 3(a). The 
bottom row in Fig. 4 shows the corresponding results for the disorder potential
arising from random scatterers. Since the doping concentration in this case 
coincides with the concentration of the scatterers, the holes are trapped at 
the impurity sites. This leaves the surrounding effectively undoped and 
thereby induces orbital ordering. This is apparent from the spread of the 
dark-blue regions and their cross correlation with the charge density 
distribution in the bottom row of Fig. 4. Such a picture with orbitally 
ordered regions coexisting with orbitally disordered patches describes 
perfectly the double peak structure of the distribution function in Fig. 3(b).

\begin{figure}
\centerline{
\includegraphics[width=8.6cm , clip=true]{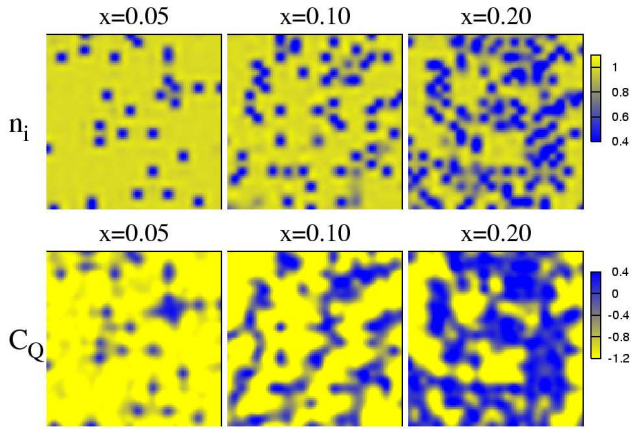}
}
\vspace{.1cm}
\caption{(Color online)
Doping evolution of the local charge density $n_i$ and the local lattice 
correlations $C_Q(i)$ for a single realization of random scatterers with
strength $D=1$, on a $24 \times 24$ lattice.
}
\end{figure}

Although we are primarily interested in the experimentally relevant case 
$x=0.3$, it is useful to see how the real-space patterns evolve as one moves 
from low to high hole densities. The undoped system is an orbitally ordered 
insulator, which turns into an orbitally disordered metal upon doping 
\cite{van-aken}. We show real-space patterns at three different doping 
concentrations in Fig. 5. The density of random scatterers is kept equal to 
the doping fraction $x$. The charge density distribution is largely controlled 
by the disorder distribution. At low doping, disconnected orbitally disordered 
regions are essentially tied to the trapped holes. With increasing $x$ the 
orbitally disordered regions begin to connect in one-dimensional snake-like 
patterns. By further increasing the doping and the concentration of scattering 
centers the orbitally disordered regions grow. The phenomenon
of disorder-induced orbital ordering is likely to be present only in a narrow doping range
near and above $0.25$, because for $x < 0.25$, the system is orbitally ordered even in the clean limit.

It is worthwhile to point to a similarity between the effects of disorder in 
the present study and in a model analysis for {\it d}-wave superconductors 
with non-magnetic impurities. In Ref. \cite{dis-ind-AFM} it was found that the 
impurities nucleate antiferromagnetism in their near vicinity. Upon increasing 
the impurity concentration static antiferromagnetism is observed. There seems 
to be a perfect analogy between the two situations, if one interchanges 
antiferromagnetism by orbital-ordering; both are ordering phenomena with the 
staggered ordering wavevector ($\pi,\pi$). The ($\pi,\pi$) ordering phenomena 
is partially triggered by the charge inhomogeneities in both cases. An 
additional complication in the present case arises from the spin degrees of 
freedom in addition to the orbital variables and from the anisotropy of the 
hopping parameters.

\begin{figure}
\centerline{
\includegraphics[width=8.8cm , clip=true]{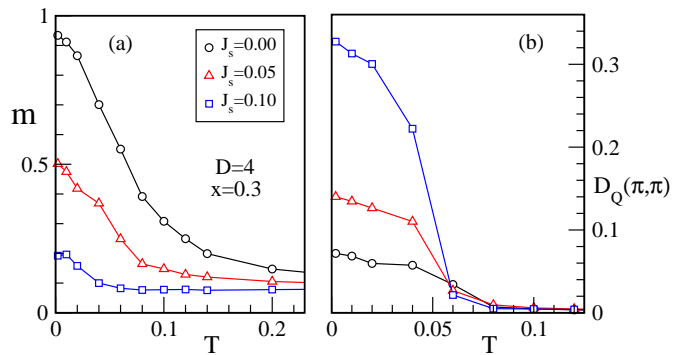}
}
\vspace{.1cm}
\caption{(Color online)
Temperature dependence of the magnetization $m$ and the staggered lattice 
structure factor $D_Q(\pi,\pi)$ for varying superexchange coupling strength 
$J_s$. The results are for random scatterers with strength $D=4$.
}
\end{figure}

As inferred above from the results for the temperature dependent magnetization
$m(T)$, the magnetic groundstate appears to be homogeneous for binary disorder,
but may be inhomogeneous in the case of doped scatterers (see Figs. 1(a),(c)). 
Since the magnetism is partially controlled by the antiferromagnetic 
superexchange coupling $J_s$, we study the effect of increasing $J_s$ for a 
fixed large disorder strength of random scatterers. Fig. 6(a) shows the result 
for $m(T)$ and Fig. 6(b) the result for the temperature dependence of 
$D_Q({\bf q}_0)$. The saturation value of $m(T)$ as well as the onset scale 
for ferromagnetism decrease with increasing $J_s$. More importantly, 
$D_Q({\bf q}_0)$ at low temperatures increases with increasing $J_s$ indicating
an enhancement in the orbital ordering. For a homogeneous system this would 
mean that orbital ordering and antiferromagnetism are both enhanced with 
increasing $J_s$. This is a contradiction to the Goodenough-Kanamori rules,
which state that an orbitally antiferro system should be magnetically ferro.
The contradiction is resolved by analyzing the microscopic details of this 
complicated state providing an example where the real-space structures are 
essential for a comprehensive understanding.

\begin{figure}
\centerline{
\includegraphics[width=8.6cm , clip=true]{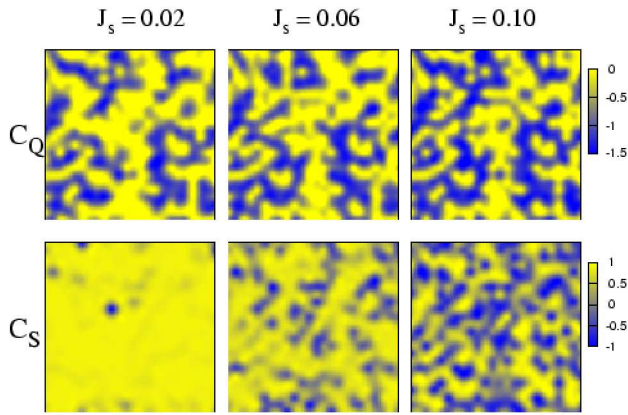}
}
\vspace{.1cm}
\caption{(Color online)
Real-space patterns for the lattice correlations $C_Q(i)$, and the analogously
defined spin correlations $C_S(i)$ for varying superexchange coupling $J_s$.
The patterns are shown for a single disorder realization at $T=0.01$. 
Orbitally ordered regions tend to maintain ferromagnetism, while the orbitally 
disordered regions are more susceptible towards antiferromagnetism with 
increasing $J_s$.
}
\end{figure}

We show in Fig. 7, the effect of the superexchange coupling on the real-space 
patterns of lattice and spin variables. The lattice correlations are shown in 
the top row and the analogously defined spin correlations $C_S(i)=(1/4)
\sum_{\delta}{\bf S}_i\cdot {\bf S}_{i+\delta}$ in the bottom row. For $J_s=
0.02$ the system contains orbitally ordered nanoscale regions, but 
magnetically it appears homogeneous. For $J_s=0.06$ the area of the orbitally 
ordered regions is enlarged and magnetic inhomogeneities appear. The 
orbitally ordered regions remain ferromagnetic, while the orbitally disordered
regions become antiferromagnetic upon increasing $J_s$. In the orbitally ordered clusters of
this inhomogeneous system in the selected parameters regime the Goodenough-Kanamori rules are therefore
fulfilled. However, upon increasing $J_s$ further to $0.1$, the antiferromagnetic regions
start to extend also into the orbitally ordered clusters.
The charge density patterns (not shown here) are insensitive to the increase 
in $J_s$.

\section{Conclusions}

Our analysis for a two-band double-exchange model for manganites leads us to conclude that
the disorder induced orbital ordering in manganites near $x=0.3$ is properly described, if
the density of scattering centers tracks the hole concentration. Within this specific model of
quenched disorder the induced staggered orbital ordering is
responsible for the orders of magnitude increase in the
low-temperature resistivity as observed in the experiments in Ref.
\cite{attfield-01}. 

\begin{center}
{\normalsize {\bf ACKNOWLEDGMENTS}}
\end{center}
SK acknowledges support by "NanoNed", a nanotechnology programme of the
Dutch Ministry of Economic Affairs.
APK gratefully acknowledges support by the 
Deutsche Forschungsgemeinschaft through SFB 484.
Simulations were performed on the Beowulf Cluster at HRI, Allahabad (India).

{}
\end{document}